\documentclass[aps,prl,reprint,superscriptaddress]{revtex4-1}
\usepackage{graphicx}
\usepackage{natbib}
\usepackage{amsmath}
\usepackage{xcolor}
\usepackage{comment}
\usepackage{soul}
\usepackage{xfrac}
\usepackage{mathtools}
\usepackage[colorlinks=true, urlcolor  = blue, citecolor = blue, linkcolor = blue]{hyperref}
\usepackage{textgreek}

\begin{document}

\title{$60^{\circ}$ and $120^{\circ}$ Domain Walls in Epitaxial BaTiO$_{3}$(111)/Co Multiferroic Heterostructures}

\author{K\'{e}vin J. A. Franke}
\affiliation{School of Physics and Astronomy, University of Leeds, Leeds LS2 9JT, United Kingdom}

\author{Colin Ophus}
\affiliation{National Center for Electron Microscopy, Molecular Foundry, Lawrence Berkeley National Laboratory, Berkeley, California 94720, USA}

\author{Andreas K. Schmid}
\affiliation{National Center for Electron Microscopy, Molecular Foundry, Lawrence Berkeley National Laboratory, Berkeley, California 94720, USA}

\author{Christopher H. Marrows}
\affiliation{School of Physics and Astronomy, University of Leeds, Leeds LS2 9JT, United Kingdom}

\date{\today}

\begin{abstract}
We report on domain pattern transfer from a ferroelectric BaTiO$_{\mathrm{3}}$ substrate with a $(111)$-orientation of the surface to an epitaxial Co film grown on a Pd buffer layer. Spatially modulated interfacial strain transfer from ferroelectric/ferroelastic domains and inverse magnetostriction in the ferromagnetic film induce stripe regions with a modulation of the in-plane uniaxial magnetic anisotropy direction. Using spin-polarized low energy electron microscopy, we observe the formation of two distinct anisotropy configurations between stripe regions. Moreover, through application of a magnetic field parallel or perpendicular to these stripes, head--to--head or head--to--tail magnetization configurations are initialized. This results in four distinct magnetic domain wall types associated with different energies and widths, which in turn affects whether domain pattern transfer can be achieved. 
\end{abstract}
\maketitle

\begin{figure}
\centering
\includegraphics{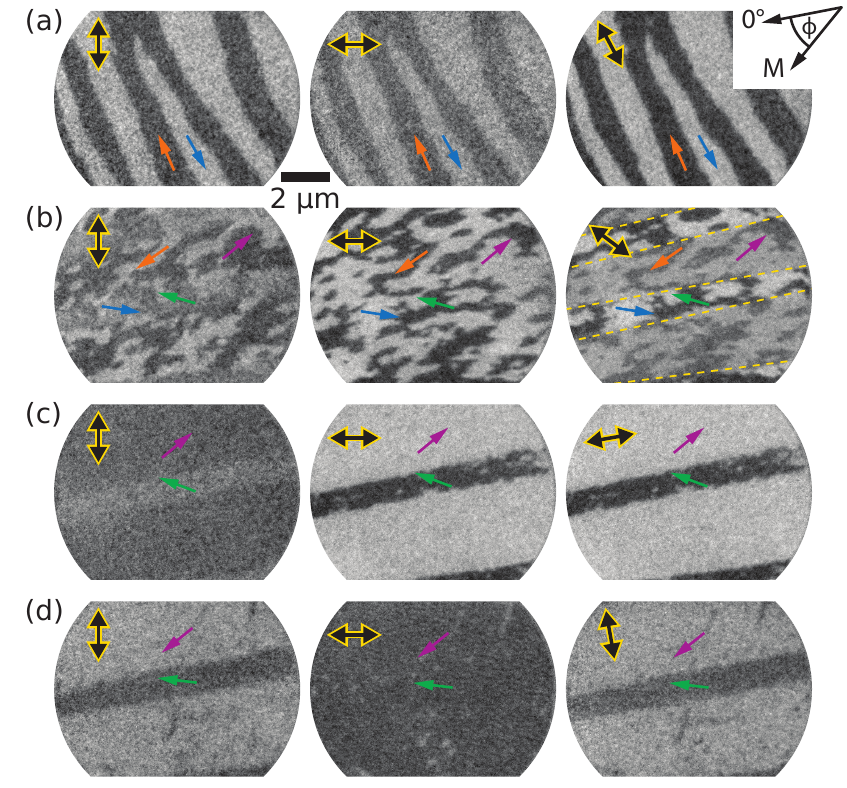}
\caption{\label{BTO_images4s5} Evolution of magnetic domain pattern and magnetization orientation in $4$ nm thick Co: (a) As deposited film. (b) After heating to $350$ $^{\circ}$C and returning to room temperature. (c) After applying a saturating magnetic field perpendicular and (d) parallel to the stripe domain orientation. Double-headed arrows indicate the axis of magnetic contrast and single-headed arrows the local orientation of the magnetization.}
\end{figure}

\begin{figure}
\centering
\includegraphics{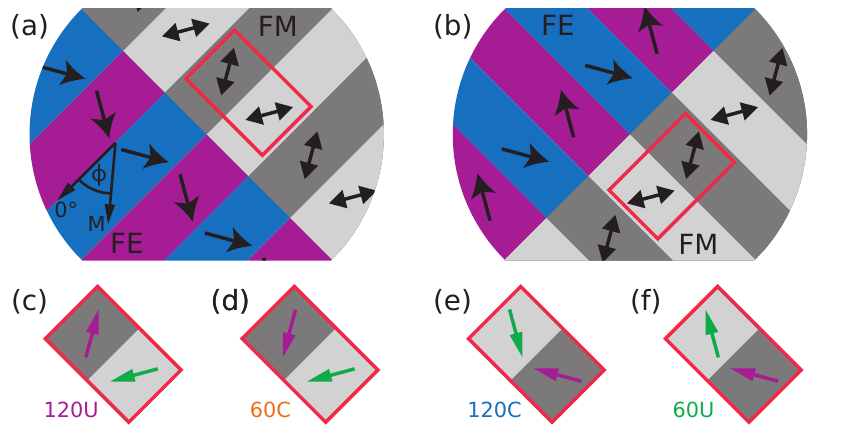}
\caption{\label{domain_sketch} Sketches of ferroelectric (color) and magnetic anisotropy (gray) domain patterns for a Co film strain coupled to a BaTiO$_{\mathrm{3}}$ $(111)$ substrate corresponding to (a) quasiparallel and (b) quasiperpendicular anisotropy patterns. (c -- f)  Possible orientations of the magnetization in the domains leading to four distinct magnetic domain walls.}
\end{figure}

\begin{figure}
\centering
\includegraphics{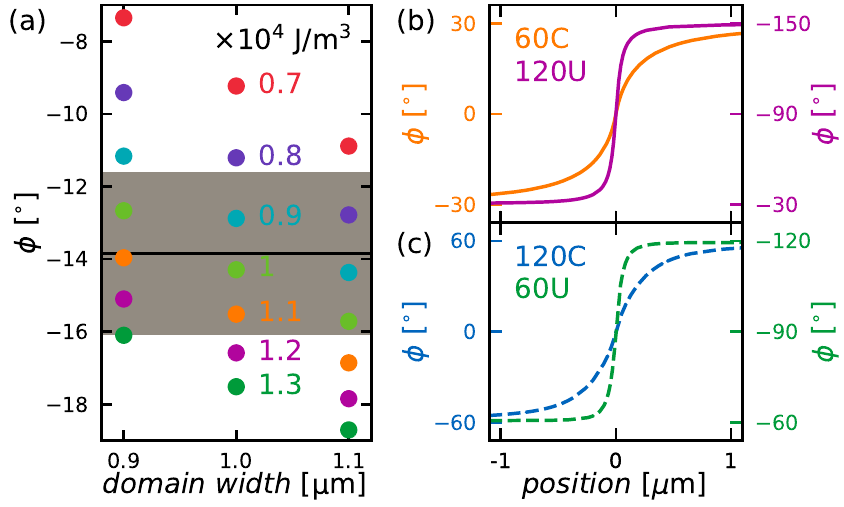}
\caption{\label{simulations} (a) Determination of the anisotropy strength from the magnetization orientation in Fig.~\ref{BTO_images4s5}(d). The width of the central domain ranges from $0.9$ to $1.1$ \textmu m.  The experimentally determined magnetization angle (black line, error in gray) is compared to the magnetization angle obtained from micromagnetic simulations for anisotropy strengths ranging from $0.7$ to $1.3 \times 10^{4}$ J/m$^{3}$ (colored dots). (b) \& (c) Magnetization profiles of all four magnetic domain wall types.}
\end{figure}

\begin{figure}
\centering
\includegraphics{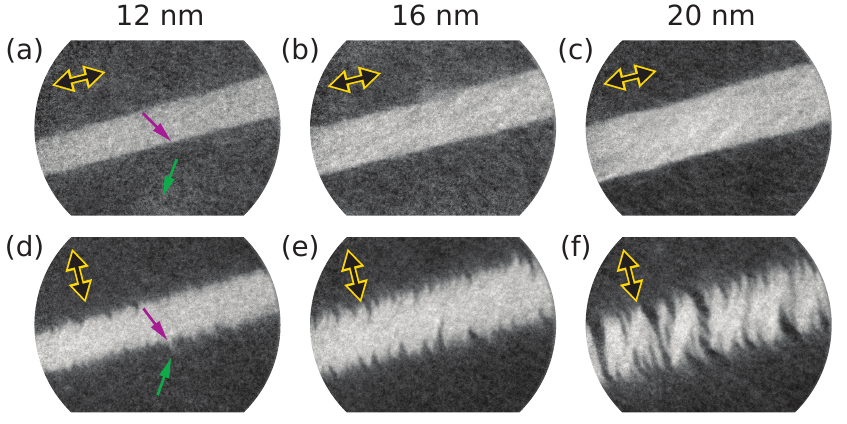}
\caption{\label{BTO_images4s6} Magnetic domain configurations for quasiperpendicular anisotropy domain wall as a function of film thickness.}
\end{figure}

Multiferroic heterostructures are often used to modify and control the properties of ferromagnetic films via interfacial coupling to a ferroelectric substrate \cite{PhysRevLett.103.257601, Chu2008, You_Origin_2013, Ghidini2015, Ghidini_JAP_2021, PhysRevLett.126.117603}. Imprinting of ferroelectric domain patterns into ferromagnetic thin films has been achieved for example via exchange coupling from $(001)$-oriented multiferroic BiFeO$_{\mathrm{3}}$ \cite{chu_electric-field_2008, trassin_interfacial_2013, PhysRevApplied.10.054030}. Alternatively, interfacial strain transfer from ferroelectric/ferroelastic domains in tetragonal $(001)$-oriented BaTiO$_{\mathrm{3}}$ induces spatially modulated magnetoelastic anisotropies in ferromagnetic thin films through inverse magnetostriction. Domain pattern transfer is achieved when the induced anisotropy overcomes intrinsic properties of the ferromagnetic film, namely the magnetocrystalline anisotropy, the exchange interaction and magnetostatics \cite{lahtinen_pattern_2011, Lahtinen_alternating_2012, streubel_strainmediated_2013, Shirahata2015}. As a result, magnetic domain walls are strongly pinned onto their ferroelectric counterparts, which has been used to drive magnetic domain wall motion with an applied voltage \cite{PhysRevX.5.011010, Diego_electric_2017}. Tuning and switching of magnetic domain wall properties, such as their width and chirality can be achieved with an applied magnetic field \cite{Diego_electric_2017, PhysRevB.85.094423,PhysRevLett.112.017201}. \\
In previous studies, domain wall coupling between $(001)$-oriented BaTiO$_{\mathrm{3}}$ substrates exhibiting $a_1$-$a_2$ domains (where the polarization rotates by $90^{\circ}$ between regular stripe domains) and in-plane magnetized thin films has been investigated extensively \cite{PhysRevB.85.094423,PhysRevLett.112.017201,Casiraghi_Influence_2015}. The magnetization in the pinned ferromagnetic domain walls rotates by $90^{\circ}$ and two distinct domain wall types can be initialized \cite{PhysRevB.85.094423}: after the application of a saturating magnetic field perpendicular to the stripes, magnetically uncharged head-to-tail domain walls are formed. In this case, the domain wall width is determined by a competition between the uniaxial magnetic anisotropy and the short range exchange interaction. \\
Magnetically charged head-to head domain walls are formed after the saturating magnetic field has been applied parallel to the stripes. The domain wall width is now determined by a competition between the magnetic anisotropy and long range magnetostatics. The width of charged domain walls has been shown to be orders of magnitude wider than the width of uncharged domain walls, and to increase linearly with film thickness (unlike uncharged domain walls) \cite{hubert_charged_1979, PhysRevLett.112.017201}.
Domain pattern transfer breaks down when the width of domain walls becomes comparable to the width of the anisotropy stripe regions. This breakdown occurs for different stripe widths and film thicknesses depending on which type of domain wall is initialized. \\
Here, we investigate the imprinting of the ferroelectric domains of $(111)$-oriented BaTiO$_{\mathrm{3}}$ substrates into epitaxial Co films through a Pd buffer layer. Using spin-polarized low energy electron microscopy (SPLEEM), we are able to investigate at which point in the sample preparation domain pattern transfer occurs. We show that four distinct domain wall types can be stabilized, and explore their effect on domain pattern transfer as a function of film thickness. These heterostructures could enable electric field tuning of magnetic domain walls.  \\
SPLEEM affords high resolution imaging of magnetic contrast combined with in situ thin film deposition enabling imaging of the magnetization direction during thin film growth \cite{Grzelakowski1996, Rougemaille2010}. Magnetic contrast is obtained because the amount of electrons backscattered elastically from the surface of a magnetic sample depends also on the relative orientation of the spin polarization of the incident electrons and the magnetization. The spin polarization of the incident electron beam can be precisely oriented in any spatial direction allowing for control over the direction of magnetic contrast. Here thin films are deposited in situ at room temperature by molecular beam epitaxy (MBE). \\ 
Figure~\ref{BTO_images4s5}(a) shows the domain pattern observed after the deposition of $4$ nm of Co.
Double-headed arrows indicate the direction of in-plane magnetic contrast, while arrows show the orientation of the magnetization in the domains. 
The observed pattern corresponds to a demagnetized domain pattern where adjacent stripe domains are separated by $180^{\circ}$ magnetic domain walls. \\
After annealing the sample at $350$ $^{\circ}$C the domain pattern shown in Figure~\ref{BTO_images4s5}(b) is observed at room temperature. The demagnetization pattern has been altered. The magnetization now aligns along uniaxial directions that rotate by $60^{\circ}$ between stripe regions, which is particularly clear in the rightmost image. This observation indicates the presence of a uniaxial anisotropy that rotates by $60^{\circ}$ between adjacent anisotropy stripe regions. \\
After saturating the magnetization perpendicular to these stripe regions the magnetization aligns with these anisotropy directions as shown in Figure~\ref{BTO_images4s5}(c). This results in a head-to-tail configuration of the magnetization in adjacent stripes. The magnetization in the wider stripes (violet arrow) makes a $29\pm2^{\circ}$ angle with the domain wall, while the magnetization in the narrow stripe (green arrow) is rotated $32\pm3^{\circ}$ away from the domain wall. The spin rotation $\Delta \phi = |\phi_1 - \phi_2|$, defined as the difference in magnetization between neighboring stripes is therefore $\Delta \phi = 119\pm4^{\circ}$. In the domain wall, the magnetization rotates through a direction perpendicular to the domain wall. \\
Figure~\ref{BTO_images4s5}(d) shows the domain pattern observed after saturating the magnetization parallel to the stripe domains. The magnetization in neighboring stripes is now oriented in a head-to-head configuration. The magnetization in the wide stripes has been rotated by exactly $180^{\circ}$, and still makes an angle of $29\pm2^{\circ}$ with the domain wall. The magnetization in the narrow central domain has not flipped, but now makes an angle of only $14\pm2^{\circ}$ with the domain wall. The spin rotation is reduced to $43\pm3^{\circ}$, and the magnetization rotates through a direction parallel to the domain wall. \\
We will now show that these results can be explained by interfacial strain transfer from the $(111)$-oriented BaTiO$_{\mathrm{3}}$ substrate after thin film deposition and inverse magnetostriction in the Co film. At room temperature BaTiO$_{\mathrm{3}}$ is in its tetragonal phase where the cubic parent phase is elongated by $1.1\%$ along a $\langle001\rangle$ direction. This elongation coincides with a ferroelectric polarization that points in the same direction. From investigations of $(001)$-oriented BaTiO$_{\mathrm{3}}$ substrates it is known that strain relaxation leads to the formation of ferroelectric stripe domains where the polarization rotates by $90^{\circ}$ between neighboring domains. When projected onto the $(111)$ surface considered here, this results in an angle of $60^{\circ}$ between the lattice elongation in adjacent stripe domains. One resulting projection of the ferroelectric polarization onto the $(111)$ surface is sketched in color in Figure~\ref{domain_sketch}(a), where the polarization rotates by $60^{\circ}$ between stripe domains. 
If the associated lattice elongation is transferred to a magnetostrictive thin film with negative magnetostriction constant $\lambda$ (as is the case for Co), it results in the uniaxial anisotropy pattern sketched in gray. Note that only the in-plane component of the elongation of the substrate lattice can impose a strain on a thin film, while the out-of-plane component cannot. The anisotropies make a $\pm30^{\circ}$ angle with the domain boundaries. For convenience we will thus refer to the anisotropy pattern sketched in Figure~\ref{domain_sketch}(a) as ``quasiparallel''.  The ``quasiperpendicular'' anisotropy pattern sketched in Figure~\ref{domain_sketch}(b) where the anisotropies make a $\pm60^{\circ}$ angle with the domain boundaries will be discussed later.\\
When the magnetization aligns with the uniaxial anisotropies of the quasiparallel configuration, two distinct types of magnetic domain wall can be formed as sketched in Figure~\ref{domain_sketch}(c) \& (d) : The magnetization can be initialized in a head-to-head (or equivalent tail-to-tail) configuration with a spin rotation of $\Delta \phi = 60^{\circ}$. This leads to an accumulation of magnetic charges and the domain wall is thus labeled ``$60$C'' for its spin rotation and the fact that it is magnetically charged. Alternatively, a head-to-tail configuration of the magnetization with a spin rotation of $\Delta \phi = 120^{\circ}$ can be formed, yielding a magnetic domain wall labeled ``$120$U'', as it does not exhibit a net magnetic charge (uncharged). These two configurations closely resemble the orientations of the magnetization observed in Figure~\ref{BTO_images4s5}(c) and (d). The only discrepancy is observed for the magnetization in the narrow central domain in the ``$60$C'' configuration, as it makes a $14\pm2^{\circ}$ angle with the domain wall, instead of the expected angle of $30^{\circ}$. \\
We investigate this reduction in angle using micromagnetic simulations and use them to determine the strength of the uniaxial anisotropy induced in the Co film. We use OOMMF \cite{donahue_oommf_1999}  to simulate a 4 nm thick magnetic film with saturation magnetization $M_{\mathrm{s}} = 1.46\times10^6$ A/m and exchange stiffness $A = 3.1\times 10^{-11}$ \cite{coey_2010}. Close inspection of the images in Figure~\ref{BTO_images4s5} reveals that the width of the central domain varies from $0.9$ to $1.1$ \textmu m. Ferroelectric domain walls are known to not be perfectly parallel. We thus simulate a 7 \textmu m wide area with a central narrow stripe surrounded by two wide stripes. A uniaxial anisotropy is considered, where the easy axis is along $\phi_{1} = -30^{\circ}$ in the narrow central domain, and along $\phi_{2} = 30^{\circ}$ in the wide domains. Two-dimensional boundary conditions are used to simulate an extended film \cite{wang_twodimensional_2010} . We vary the width of the narrow domain and the strength of the uniaxial anisotropy. \\
In Figure~\ref{simulations}(a) the resulting magnetization angle in the central domain for the head-to-head magnetization configuration is compared to the experimental observation. 
From this comparison, we find a match of the magnetization angle for an uniaxial anisotropy constant $K_{\mathrm{u}} =1\pm0.2 \times 10^4$ J/m$^3$. For 120U domain walls the magnetization in the central domain aligns with the uniaxial anisotropy axis along $\phi_{1} = -30^{\circ}$ both for the experimental observation and the micromagnetic simulations in the anisotropy range considered here (not shown). \\
We now turn our attention to the origin of this reduction in spin rotation. As the magnetization in wide domains is seen to align with the uniaxial anisotropy direction, we repeat the micromagnetic simulations for an anisotropy constant $K_{\mathrm{u}} =1 \times 10^4$ J/m$^3$, but now for anisotropy stripes of equal width of  $4$ \textmu m. We show the resulting magnetic domain wall profiles in Figure~\ref{simulations}(b). We observe that despite having a smaller spin rotation, the accumulation of magnetic charges in the $60$C magnetic domain wall leads to a more gradual rotation of the magnetization, and therefore a wider domain wall, than for the $120$U magnetic domain wall. Even $1$ \textmu m from the magnetic domain wall center the magnetization of the $60$C magnetic domain wall has not fully aligned with the anisotropy axis, yet. As the domain walls in Figure~\ref{BTO_images4s5}(d) are only $1 \pm 0.1$ \textmu m apart, the magnetization can never fully align with the anisotropy axis, because the tails of domain walls overlap. \\
For the ferroelectric polarization orientations sketched in Figure~\ref{domain_sketch}(a), the domain configuration sketched in Figure~\ref{domain_sketch}(b) is in principle energetically equivalent. When the in-plane lattice elongation associated with this ferroelectric/ferroelastic domain pattern is transferred to the Co film, inverse magnetostriction is expected to induce the anisotropy pattern schematically shown in gray, and labeled ``quasiperpendicular''. It is distinct from the quasiparallel anisotropy stripe pattern, as can easily be seen from the type of magnetic domain walls that are obtained when the magnetization aligns with the anisotropy directions: after saturating the magnetization parallel to the domain walls, ``$120$C'' magnetic domain walls are obtained, whereas ``$60$U'' magnetic domain walls are initialized after saturation perpendicular to the domain boundaries [Figure~\ref{domain_sketch}(e) \& (f)]. The angular profiles of these domain walls are shown in Figure~\ref{simulations}(c), confirming that the width of domain walls depends significantly on the magnetostatic energy, and less on the spin rotation. \\
We show experimental evidence for these domain walls in Figure~\ref{BTO_images4s6}.
These images were taken on a second sample that was prepared in the same manner as the one imaged in Figure~\ref{BTO_images4s5}, except that a $12$ nm thick Co film was grown. We observe that after annealing and the application of a saturating magnetic field in the appropriate direction, a ``$60$U'' magnetic domain wall is imaged in Figure~\ref{BTO_images4s6}(a), while a ``$120$C'' magnetic domain wall is depicted in Figure~\ref{BTO_images4s6}(d). We have therefore observed all four possible magnetic domain wall types sketched in Figure~\ref{domain_sketch}. \\
We continue by investigating the film thickness dependence of the magnetic domain configurations for quasiperpendicular anisotropy domain walls. As the SPLEEM instrument affords in-situ  thin film deposition, the film thickness can easily be increased and images taken at the same position. After each additional thin film deposition the sample is annealed to ensure strain transfer throughout the thin film. The ferroelectric domain pattern is expected to be reinitialized during the annealing process, but ferroelectric domain walls tend to be pinned by defects, and strains (imposed by the holder) and the shape of the substrate are all expected to have an influence on the ferroelectric domain pattern and location of ferroelectric domain walls \cite{Forsbergh1949, Little1955, MUNOZSALDANA2001L402, Lukyanchuk2009, tagantsev2010domains}. Fortunately, as seen in Figures~\ref{BTO_images4s6}(b) \& (c), the domain pattern does not change significantly after increasing the film thickness to $16$ nm and $20$ nm. A widening of the central domain is observed. \\
Note that the ``$60$U'' magnetic domain walls observed in Panels (a) -- (c) are sharp and straight. In contrast, the charged domain wall depicted in Panels (d) -- (f) is a zig-zag wall, thus avoiding the accumulation of magnetic charges. At $20$ nm [Panel(f)], this even leads to the formation of magnetic stripe domains within the central anisotropy stripe. As described before, such a behavior is not observed for ``$60$U'' magnetic domain walls. Clearly, the energy cost of forming a charged magnetic domain wall over an uncharged one increases with film thickness, as expected. \\
We again investigate this further using micromagnetic simulations of $2.5$ \textmu m wide anisotropy stripes of equal width. Note that the width of the central stripe in Figure~\ref{BTO_images4s6} ranges from $1.5$ \textmu m at $t=12$ nm to $2.5$ \textmu m at $t=20$ nm.
\begin{figure}
\centering
\includegraphics{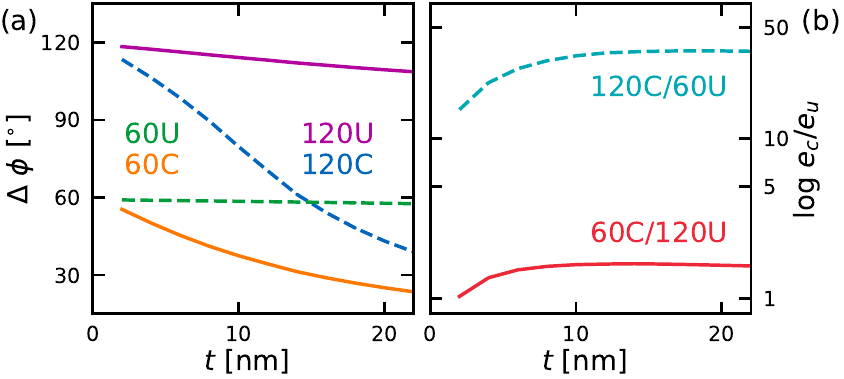}
\caption{\label{t-dep_sims_time} Dependence of (a) the spin rotation $\Delta \phi$, and (b) the ratio between energy densities, as a function of magnetic film thickness $t$ for $2.5$ \textmu m wide anisotropy stripes.}
\end{figure}
The use of two-dimensional periodic boundary conditions and the fact that simulations do not take into account defects in the film precludes the formation of magnetic domains within anisotropy stripes. This makes it easier to compare the different types of magnetic domain walls. \\
Figure~\ref{t-dep_sims_time}(a) shows the spin rotation $\Delta \phi$ as a function of thickness for all four domain wall types. While little change is observed for both types of uncharged domain walls, charged domain walls exhibit a clear reduction of their spin rotation when the film thickness is increased. The general trend can be understood from the expected thickness dependence of domain wall widths. While the width of uncharged domain walls is expected to be independent of the film thickness, the accumulation of a net magnetic charge at the domain wall leads to a linear dependence of the domain wall width on the film thickness. In the latter case, an increase of the film thickness therefore results in a larger overlap of neighboring domain walls, and in turn to a reduction in spin rotation. \\
Figure~\ref{t-dep_sims_time}(b) shows the ratio between energy densities of domain patterns with charged and uncharged magnetic domain walls. We see that for $2$ nm thick films, the 120C configuration that we attempted to initialize in Figure~\ref{BTO_images4s6}(d) is $15$ times higher in energy than the corresponding uncharged (60U) domain configuration. For a $20$ nm thick film this ratio increases to $35$, making the formation of 120C magnetic domain walls energetically very costly. As a result, we observe the roughening of magnetic domain walls for thin magnetic films in Figure~\ref{BTO_images4s6}(d) and the breakdown of pattern transfer and fromation of magnetic domains within anisotropy stripe domains for thicker films [Figure~\ref{BTO_images4s6}(f)]. In comparison, the energy cost of forming 60C domain walls over 120U ones in a quasiparallel anisotropy pattern (depicted in Figure~\ref{BTO_images4s5}) is much lower. \\
So far, we have attributed the formation of magnetic stripe domains observed in Figures~\ref{BTO_images4s5} and \ref{BTO_images4s6} to coupling of the ferromagnetic thin films to the $(111)$-oriented BaTiO$_{\mathrm{3}}$ substrates via interfacial strain transfer and inverse magnetostriction.
\begin{figure}
\centering
\includegraphics{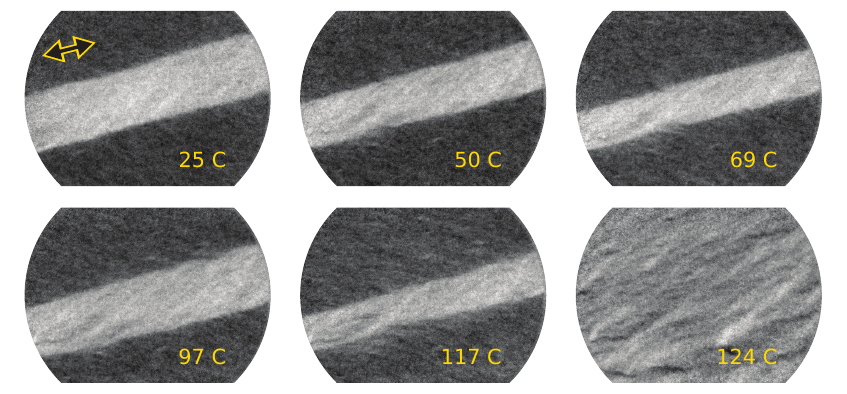}
\caption{\label{BTO_4s6_images_50ML_temp} Evolution of 60U domain pattern as a function of temperature.}
\end{figure}
While our data strongly support this interpretation, we are unable to image the ferroelectric BaTiO$_{\mathrm{3}}$ domains in SPLEEM, and are thus unable to definitely prove a one-to-one correlation of ferroelectric and ferromagnetic domains. We are however able to provide further evidence for our claim. \\
As described earlier, when heated BaTiO$_{\mathrm{3}}$ undergoes a transition from a tetragonal to a cubic phase at around $120$ $^{\circ}$C \cite{KayVousden1949}. The strain imposed on the ferromagnetic thin film and the induced uniaxial anisotropy are thus expected to vanish at the phase transition. Figure~\ref{BTO_4s6_images_50ML_temp} shows a series of SPLEEM images taken as a function of temperature. The first image is the same as the one in Figure~\ref{BTO_images4s6}(c). We are thus imaging a ``$60$U'' magnetic domain wall configuration in a $20$ nm thin Co film. Images are taken in exactly the same location. The change in width of the central domain as a function of temperature is therefore not an artifact. It seems that the upper domain wall moves with temperature. We can only speculate about the origin of this movement, but a temperature activated motion of ferroelectric domain walls between local pinning sites or a change in internal stresses of the constrained substrate seem likely \cite{Mitsui1953, tagantsev2010domains}. Most importantly, the magnetic stripe pattern disappears between the images collected at $117^{\circ}$ C and $124^{\circ}$ C, exactly where the ferroelectric phase transition of BaTiO$_{\mathrm{3}}$ occurs. These images therefore confirm that magnetic stripe domains are formed via coupling to the substrate in its tetragonal phase. \\
We now discuss how this new heterostructure holds the potential of unprecedented electric field tuning of magnetic domain walls. So far, the application of an electric field along the $[001]$-direction of a BaTiO$_{\mathrm{3}}$ substrate exhibiting $a_1$-$a_2$ domains eventually yields a switch of the polarization and associated lattice elongation towards the out-of-plane direction. As a result the spatial modulation of strain that affords domain pattern transfer to a ferromagnetic thin film is lost \cite{Franke2016ch21}. \\
In $(111)$-oriented BaTiO$_{\mathrm{3}}$ substrates the polarization makes a $\approx 54.7^{\circ}$ angle with the surface normal. In the tetragonal phase, the application of an electric field in the out-of-plane direction cannot switch the polarization into the $[111]$-direction. Instead, the lattice is deformed, thus modulating the strain imposed onto the ferromagnetic film and therefore altering the magnitude of the anisotropy constant. This is turn affects the width of magnetic domain walls. Moreover, the application of large electric fields yields a transition to an orthorombic phase, with the same symmetry of induced anisotropy directions, but reduced anisotropy constant. Finally, a second transition to rhombohedral phase at even larger fields aligns the polarization with the electric field direction, and spatial modulations of strain are erased \cite{Wada1999, Bell2001}. \\
In summary, we have demonstrated domain pattern transfer from $(111)$-oriented BaTiO$_{\mathrm{3}}$ substrates to epitaxial Co films via spatially modulated interfacial strain transfer and inverse magnetostriction. Two types of anisotropy patterns can be induced in the ferromagnetic film, and magnetic fields can be used to initialize magnetically charged or uncharged domain walls with a spin rotation of $60^{\circ}$ or $120^{\circ}$. In our samples, pattern transfer occurs only after annealing through the ferroelectric phase transition, indicating strain relaxation in the Pd buffer layer. Our results open up new degrees of freedom for designing low-power spintronic devices. We moreover expect that these multiferroic heterostructures can be used for electric field tuning of magnetic domain walls and pattern transfer. \\


\begin{acknowledgments}
This project has received funding from the European Union's Horizon 2020 research and innovation programme under the Marie Sklodowska-Curie grant agreement No 750147. K.J.A.F. acknowledges support from the Jane and Aatos Erkko Foundation. Work at the Molecular Foundry was supported by the Office of Science, Office of Basic Energy Sciences, of the U.S. Department of Energy under Contract No. DE-AC02-05CH11231. This research used the Lawrencium computational cluster resource provided by the IT Division at the Lawrence Berkeley National Laboratory (Supported by the Director, Office of Science, Office of Basic Energy Sciences, of the U.S. Department of Energy under Contract No. DE-AC02-05CH11231.
\end{acknowledgments}


%

\end{document}